\documentclass[letterpaper,english,reprint, aps]{revtex4-1}
\usepackage[T1]{fontenc}
\usepackage[latin9]{inputenc}
\usepackage{array}
\usepackage{multirow}
\usepackage{amsmath}
\usepackage{amssymb}
\usepackage{graphicx}

\makeatletter

\newcommand{\noun}[1]{\textsc{#1}}
\newcommand{\lyxmathsym}[1]{\ifmmode\begingroup\def\b@ld{bold}
  \text{\ifx\math@version\b@ld\bfseries\fi#1}\endgroup\else#1\fi}

\providecommand{\tabularnewline}{\\}

\makeatother

\usepackage{babel}
\begin{document}
\title{Investigation of Charged Higgs Boson in the Bottom and Top Quark Decay
Channel at the FCC-hh}
\author{I. Turk Cakir}
\email{ilkay.turk.cakir@cern.ch (I. Turk Cakir)}

\affiliation{Department of Energy Systems Engineering, Giresun University, 28200
Giresun, Turkey}
\author{O. Cakir}
\email{ocakir@science.ankara.edu.tr (O. Cakir)}

\affiliation{Department of Physics, Ankara University, 06100 Ankara, Turkey}
\author{H. Denizli }
\email{denizli_h@ibu.edu.tr (H. Denizli)}

\affiliation{Department of Physics, Bolu Abant Izzet Baysal University, 14280,
Bolu, Turkey}
\author{A. Senol}
\email{senol_a@ibu.edu.tr (A. Senol)}

\affiliation{Department of Physics, Bolu Abant Izzet Baysal University, 14280,
Bolu, Turkey}
\author{A. Yilmaz}
\email{aliyilmaz@giresun.edu.tr (A. Yilmaz)}

\affiliation{Department of Electrical and Electronics Engineering, Giresun University,
28200 Giresun, Turkey}
\date{\today}
\begin{abstract}
After the recent discovery of a neutral Higgs boson with a mass about
$125$ GeV, we assess the extend of discovery potential of future
circular hadron collider (FCC-hh) for a charged Higgs boson in the
bottom and top quark decay channel. The charged Higgs boson can be
produced through the $pp\to h^{-}t+X$ process with a subsequent decay
$h^{-}\to b\bar{t}$ channel. This decay channel is particularly important
for studying the charged Higgs boson heavier than the top quark. We
consider an extension of the standard model Higgs sector, namely two
Higgs doublet model (2HDM), and perform a dedicated signal significance
analysis to test this channel for the FCC-hh running at the center
of mass energy of $100$ TeV and the integrated luminosity of $1$
ab$^{-1}$ (initial) and $30$ ab$^{-1}$ (ultimate). We find that
an important part of the parameter spaces of two Higgs doublet model
are examinable at the FCC-hh.
\end{abstract}
\maketitle

\section{Introduction}

The Higgs boson have been discovered by the ATLAS \citep{key-1} and
CMS \citep{key-2} experiments at the CERN LHC in 2012. This discovery
has motivated a lot of measurements to identify the nature of the
discovered particle. We have elementary fermions (quarks and leptons)
and bosons (vectors and scalar) within the standard model (SM) of
particle physics. However, multiple scalars are predicted by some
extensions of the standard model, such as two Higgs doublet model
(2HDM) \citep{key-3,key-4}, and supersymmetry (SUSY) \citep{key-5}
(and references therein), to deal with some issues such as dark matter,
hierarchy, etc. In addition to neutral scalars, one can expect singly
or doubly charged Higgs bosons in such models. Recently, charged Higgs
boson discovery prospects have been studied in Ref. \citep{key-6},
which classify models into categories of different coupling properties.

At a center of mass energy of 13 TeV in proton proton collisions,
the ATLAS and CMS Collaborations have performed several searches for
charged Higgs bosons \citep{key-7,key-8}, where low values of $\tan\beta<1$
are excluded for a charged Higgs boson mass up to 160 GeV. The most
stringent upper limit from ATLAS on $\sigma(pp\to h^{+}t+X)\times{\cal B}(h^{+}\to\tau^{+}\nu)$
and $\sigma(pp\to h^{+}t+X)\times{\cal B}(h^{+}\to t\bar{b})$ at
95\% CL is in the range $4.2\lyxmathsym{\textendash}0.0025$ pb and
$9.6\lyxmathsym{\textendash}0.01$ pb for a charged Higgs boson mass
in the range $90-2000$ GeV \citep{key-7} and $200\lyxmathsym{\textendash}3000$
GeV \citep{key-9}, respectively.

We study charged Higgs boson at the future circular hadron collider
(FCC-hh) with center of mass energy of $100$ TeV \citep{key-10}.
We concentrate on the 2HDM model type-II scenario or MSSM scenario.
In the second section we mention about signal process as well as corresponding
SM backgrounds. Event selection via objects in the final state have
been performed over the signal and background samples within the FCC
software (FCCSW) \citep{key-11}. The cross sections for the process
$pp\to h^{-}t+X$ have been calculated for different model parameters.
Kinematic distributions of final state objects and cut flows presented
in the next section. Reconstruction of charged Higgs boson and its
invariant mass distribution is given in the third section. Finally,
statistical significance of the signal have been calculated depending
on the parameter space (mass and couplings) of the model framework.
Finally, we draw a conclusion on the search for charged Higgs boson
at the FCC-hh.

\section{Signal And Background}

For the signal, we use the scalar potential and the Yukawa sector
of the general 2HDM \citep{key-3}, in which the complex (pseudo)
scalar doublets $\Phi_{j}(j=1,2)$ can be parametrized as

\begin{equation}
\Phi_{j}(x)=\left(\begin{array}{c}
\phi_{j}^{+}(x)\\
\frac{1}{\sqrt{2}}[v_{j}+\phi_{j}^{0}(x)+iG_{j}(x)]
\end{array}\right)
\end{equation}
where $v_{1,2}$ are vacuum expectation values of two Higgs doublets
satisfying $v=\sqrt{v_{1}^{2}+v_{2}^{2}}$ with $v\simeq246$ GeV.
The ratio of the vacuum expectation values is defined $v_{1}/v_{2}=\tan\beta$
as a free parameter. Two CP-even physical field can be written in
terms of two neutral scalar fields

\begin{equation}
\left(\begin{array}{c}
H^{0}\\
h^{0}
\end{array}\right)=\left(\begin{array}{cc}
\cos\alpha & \sin\alpha\\
-\sin\alpha & \cos\alpha
\end{array}\right)\left(\begin{array}{c}
\phi_{1}^{0}\\
\phi_{2}^{0}
\end{array}\right)
\end{equation}
and the CP-odd neutral field $A^{0}=-G_{1}\sin\beta+G_{2}\cos\beta$
and charged field  $h^{\pm}=-\phi_{1}^{\pm}\sin\beta+\phi_{2}^{\pm}\cos\beta$.

After electroweak symmetry breaking, five degrees of freedom become
physical Higgs bosons (three neutral $h^{0},H^{0},A^{0}$ and two
charged $h^{+},h^{-}$), while three degrees of freedom kept by Goldstone
bosons (neutral $G^{0}$ and charged $G^{+},G^{-}$) to attribute
massive longitudinal component of gauge fields (corresponding to neutral
$Z^{0}$ and charged $W^{+},W^{-}$ bosons). The other independent
parameters are the masses ($m_{h^{0}},m_{H^{0}},m_{A},m_{h^{\pm}}$)
of the physical Higgs bosons in the alignment limit.

The cross section for signal process $pp\rightarrow h^{-}t+X$ can
be calculated at leading order integrating over parton distribution
functions through the subprocess $gb\to h^{-}t$ partonic cross section. 

\begin{align}
\sigma(pp & \to h^{-}t+X)=\int_{x_{1min}}^{x_{1max}}\int_{x_{2min}}^{x_{2max}}\hat{\sigma}(gb\to h^{-}t)\nonumber \\
 & \times dx_{1}dx_{2}f_{1}(x_{1},\mu_{F})f_{2}(x_{2},\mu_{F})
\end{align}
where $f_{i}(x_{i},\mu_{F})$ are parton distribution functions inside
each proton (hadron momentum $h_{i}$) with the parton momentum ($p_{i}$)
fractions $x_{i}=p_{i}/h_{i}$. The limits of the integrals are defined
as $x_{imax}=1$ and $(x_{1}x_{2})_{min}=\tau_{min}=\hat{s}_{min}/s$
(where $\sqrt{s}$ is the process center of mass energy of FCC-hh
taken as $100$ TeV). The partonic cross section for the subprocess
$\hat{\sigma}(gb\to h^{-}t)$ can be calculated from the process kinematics
and the matrix elements. The matrix element squared expressions ($M_{2\to2}$)
averaged over initial state (spins, colors) and summed over final
state (spins, colors) for ($2\to2$) subprocess $g(p_{1})b(p_{2})\to h^{-}(p_{3})t(p_{4})$
is given by

\begin{align}
 & <|M_{2\to2}|^{2}>=\frac{1}{24}\frac{g_{e}^{2}g_{s}^{2}|V_{tb}|^{2}}{s_{W}^{2}\tan^{2}\beta m_{W}^{2}}\nonumber \\
\times & \left\{ \frac{A_{1}(\hat{s},\hat{t},m_{h^{-}})+A_{2}(\hat{s})\tan^{2}\beta}{(\hat{s}-m_{b}^{2})^{2}}\right.\nonumber \\
 & +\frac{A_{3}(\hat{s},\hat{t},m_{h^{-}})\tan^{4}\beta}{(\hat{s}-m_{b}^{2})^{2}}\nonumber \\
 & +\frac{A_{4}(\hat{s},\hat{t},m_{h^{-}})+A_{5}(\hat{t},m_{h^{-}})\tan^{2}\beta}{(\hat{s}-m_{b}^{2})(\hat{u}-m_{t}^{2})}\nonumber \\
 & +\frac{A_{6}(\hat{s},\hat{t},m_{h^{-}})\tan^{4}\beta}{(\hat{s}-m_{b}^{2})(\hat{u}-m_{t}^{2})}\nonumber \\
 & +\frac{A_{7}(\hat{s},\hat{t},m_{h^{-}})+A_{8}(\hat{s},\hat{t},m_{h^{-}})\tan^{2}\beta}{(\hat{u}-m_{t}^{2})^{2}}\nonumber \\
 & +\left.\frac{A_{9}(\hat{s},\hat{t},m_{h^{-}})\tan^{4}\beta}{(\hat{u}-m_{t}^{2})^{2}}\right\} \label{eq:4}
\end{align}
where the Mandelstam variables $\hat{s}=(p_{1}+p_{2})^{2}$, $\hat{t}=(p_{1}-p_{3})^{2}$
and $\hat{u}=(p_{1}-p_{4})^{2}=-\hat{s}-\hat{t}+m_{b}^{2}+m_{t}^{2}+m_{h^{-}}^{2}$
are used to shorten the amplitude of the signal subprocess in Lorentz
invariant form. The coefficients $A_{i}(\hat{s},\hat{t},m_{h^{-}})$
are written in terms of the variables and they are given in detail
in the Appendix. In Eq. \ref{eq:4}, the terms with coefficients $A_{2}(\hat{s})$,
$A_{5}(\hat{t},m_{h^{-}})$, $A_{8}(\hat{s},\hat{t},m_{h^{-}})$ are
independent of $\tan\beta$, while the others can be written in terms
of a function depending on $\tan\beta$. When the mass of $b$-quark
is neglected the matrix element squared expression will depend on
$1/\tan^{2}\beta$ as explained in the Appendix.

The matrix element squared expression ($M_{1\to2}$) for decay process
($h^{-}\to b\bar{t}$) is given by 

\begin{align}
 & <|M_{1\to2}|^{2}>=\frac{3}{2}\frac{g_{e}^{2}|V_{tb}|^{2}}{s_{W}^{2}m_{W}^{2}\tan^{2}\beta}\nonumber \\
 & \times\left[m_{t}^{2}m_{h^{-}}^{2}-m_{b}^{4}\tan^{4}\beta-m_{t}^{4}\right.\nonumber \\
 & \left.+m_{b}^{2}(m_{h^{-}}^{2}\tan^{4}\beta-m_{t}^{2}(1+4\tan^{2}\beta+\tan^{4}\beta))\right]\label{eq:5}
\end{align}
where $g_{e}$ and $g_{s}$ are the electromagnetic coupling and strong
coupling corresponding to $U(1)_{Q}$ and $SU(3)_{C}$ gauge groups,
the $s_{W}$ is the sinus of Weinberg weak mixing angle $\theta_{W}$,
and $V_{tb}$ is the relevant CKM matrix element. From the expressions
Eq. (\ref{eq:4}) and (\ref{eq:5}), we obtain the $\tan\beta$ dependence
of the cross section and decay width calculations. We present decay
width $\Gamma(h^{-}\to b\bar{t})$ values in Table \ref{tab:tab1},
depending on the parameter $\tan\beta$ and charged Higgs boson mass
$500$ GeV, $1000$ GeV and $2000$ GeV as benchmark points.

\begin{table}
\caption{Decay width for process $h^{-}\to b\bar{t}$ depending on parameter
$\tan\beta$ values and different mass values. \label{tab:tab1}}

\small%
\begin{tabular}{|c|c|c|c|c|}
\hline 
$\Gamma$ (GeV) & $\tan\beta=1$ & $\tan\beta=7$ & $\tan\beta=10$ & $\tan\beta=30$\tabularnewline
\hline 
\hline 
$m_{h^{-}}=500$ GeV & $23.46$ & $0.8697$ & $1.037$ & $7.281$\tabularnewline
\hline 
$m_{h^{-}}=1000$ GeV & $56.93$ & $2.118$ & $2.524$ & $17.67$\tabularnewline
\hline 
$m_{h^{-}}=2000$ GeV & $119.20$ & $4.437$ & $5.286$ & $36.99$\tabularnewline
\hline 
\end{tabular}
\end{table}

\begin{figure}
\includegraphics[scale=0.38]{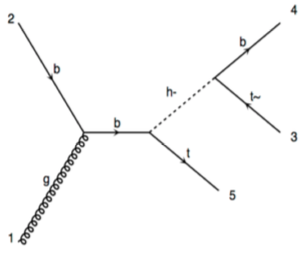}\includegraphics[scale=0.38]{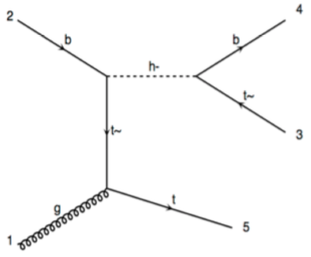}

\caption{Feynman diagrams for subprocess $gb\to h^{-}t\to b\bar{t}t$. \label{fig:fig1}}
\end{figure}

We generate the signal samples of the process $pp\to h^{-}t+X$ followed
by the decay mode $h^{-}\to b\bar{t}$ leading to an intermediate
state of a pair of top quarks and a $b$-quark. We use Pythia 8 package
\citep{key-12} for the signal event generation, where the subprocess
$gb\rightarrow h^{-}t$ already exists in this publicly available
software. The respective Feynman diagrams for the signal process are
presented in Fig. \ref{fig:fig1}. 

The decay chain, in general, ends with three possible channels depending
on the decay channels of a pair of $W$ bosons: (i) all hadronic mode
($7$ jets: 4 light jets and 3 $b$-jets), (ii) single lepton mode
(1 charged lepton and missing transverse energy, 2 light jets and
3 $b$-jets), (iii) dilepton mode (2 oppositly charged leptons and
missing transverse energy, 3 $b$-jets). We may generalize the final
state by including different type of fermion particles ($f_{i}$)
and $b$-jets ($b_{i}$) such as $f_{1}f_{2}f_{3}f_{4}b_{1}b_{2}b_{3}$.
Here, we focus on the final state including single lepton mode of
the signal: 1 lepton + MET + 3 $b$-jet + 2 jets.

Signal events are generated with Pythia 8 within the FCC software
(FCCSW) \citep{key-11} for different model parameters: mass ($m_{h^{-}}\equiv m_{H^{0}}$)
in the range of ($500-2000$) GeV, ratio of the vacuum expectation
values ($\tan\beta$) in the range of ($1-30$), and a parameter $\cos(\beta-\alpha)=0$
(alignment limit) which is relevant for $H^{0}VV$, $h^{0}AZ$ and
$h^{0}h^{\pm}W^{\mp}$ couplings. However, the background Les Houches
events (LHE) are generated with MadGraph 5 \citep{key-13}. For further
hadronization and showering for signal and background events are performed
through Pythia 8 within this software. A fast detector simulation
is performed with Delphes 3 \citep{key-14} for parametric card (FCChh.tcl)
of an FCC-hh detector. Event selection is applied on those samples
with Heppy \citep{key-15}. Flat ntuples are produced with observables
of interest and analyzed with Heppy. It reads events in FCC EDM format,
and creates lists of objects adapted to an analysis in python. The
gen-level and reco-level plots are produced with python scripts where
Heppy writes a Root program \citep{key-16} tree.

Background samples for the processes $pp\to t\bar{t}$, $pp\to t\bar{t}b$
and $pp\to t\bar{t}j$ are simulated using Delphes 3 with FCC-hh detector
card. The main background is $t\bar{t}+jets$, in particular $t\bar{t}+bjet$
in the most signal-sensitive regions.

\section{Analysis And Results}

For the signal cross section calculation we have performed benchmarking
of the parameter space of the model considered here, requiring the
mass $m_{h^{-}}$ to lie in the $500$ GeV$\lyxmathsym{\textendash}2000$
GeV range. We find signal cross sections (from Pythia 8 with generator
level defaults) as shown in Table \ref{tab:tab2}, by taking $\tan\beta$
variable and setting $\cos(\beta-\alpha)=0$. The bottom rows of Table
\ref{tab:tab2} show the cross sections for relevant SM backgrounds
obtained using MadGraph 5.

\begin{table}
\caption{Signal cross sections (in pb) depending on different mass values and
parameter $\tan\beta$, and the relevant background cross sections
at FCC-hh. \label{tab:tab2}}

\tiny%
\begin{tabular}{|c|c|c|c|c|}
\hline 
Cross sections (pb) & $\tan\beta=1$ & $\tan\beta=7$ & $\tan\beta=10$ & $\tan\beta=30$\tabularnewline
\hline 
$m_{h^{-}}=500$ GeV & $5.495\times10^{1}$ & $1.837\times10^{0}$ & $2.027\times10^{0}$ & $1.344\times10^{1}$\tabularnewline
\hline 
$m_{h^{-}}=1000$ GeV & $8.129\times10^{0}$ & $2.728\times10^{-1}$ & $2.981\times10^{-1}$ & $1.934\times10^{0}$\tabularnewline
\hline 
$m_{h^{-}}=2000$ GeV & $7.634\times10^{-1}$ & $2.558\times10^{-2}$ & $2.795\times10^{-2}$ & $1.778\times10^{-1}$\tabularnewline
\hline 
\hline 
\multicolumn{2}{|c|}{Process} & $pp\to t\bar{t}$ & $pp\to t\bar{t}j$ & $pp\to t\bar{t}b$\tabularnewline
\hline 
\multicolumn{2}{|c|}{Background cross sections (pb)} & $2.607\times10^{4}$ & $4.037\times10^{4}$ & $4.906\times10^{2}$\tabularnewline
\hline 
\end{tabular}
\end{table}

\begin{figure}
\includegraphics[scale=0.32]{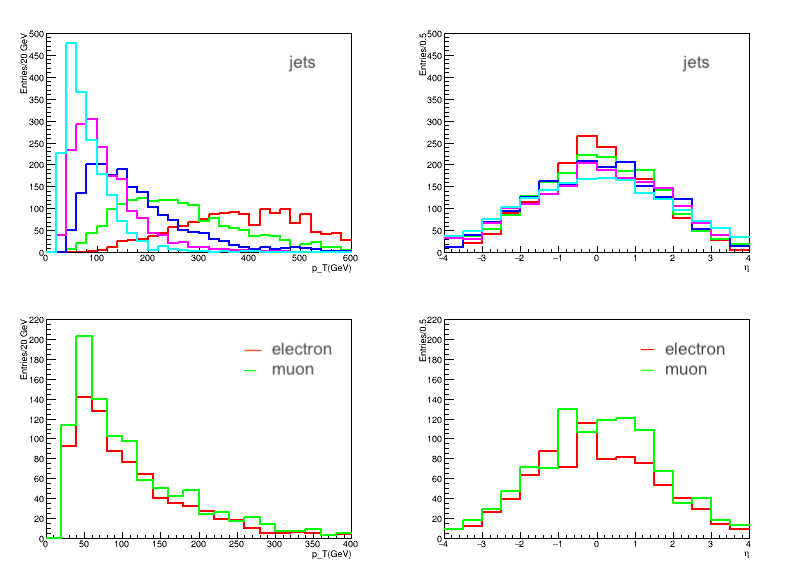}

\caption{Transverse momentum and rapidity distributions of final state detectable
objects (jets, electron or muon) for signal (mass $m_{h^{-}}=1000$
GeV). \label{fig:fig2}}
\end{figure}

\begin{figure}
\includegraphics[scale=0.32]{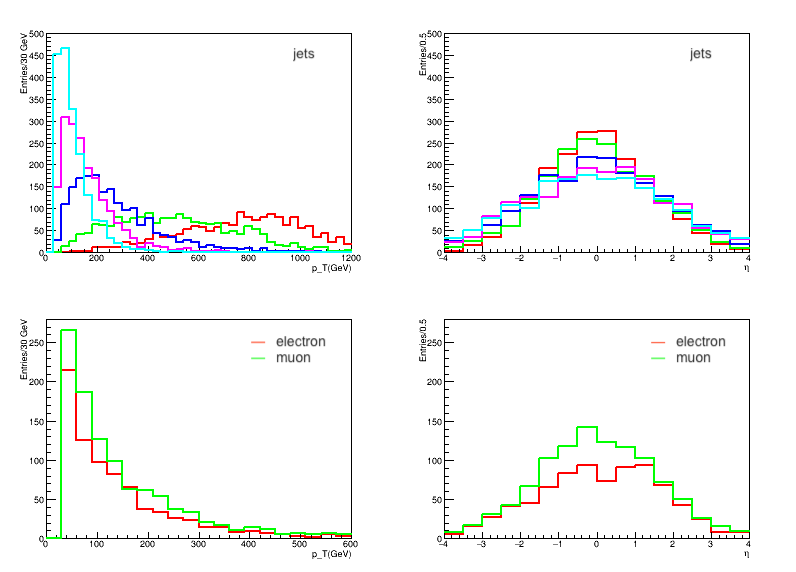}

\caption{The same as Fig. \ref{fig:fig2}, but for the charged Higgs boson
mass $m_{h^{-}}=2000$ GeV. \label{fig:fig3}}
\end{figure}

\begin{figure}
\includegraphics[scale=0.32]{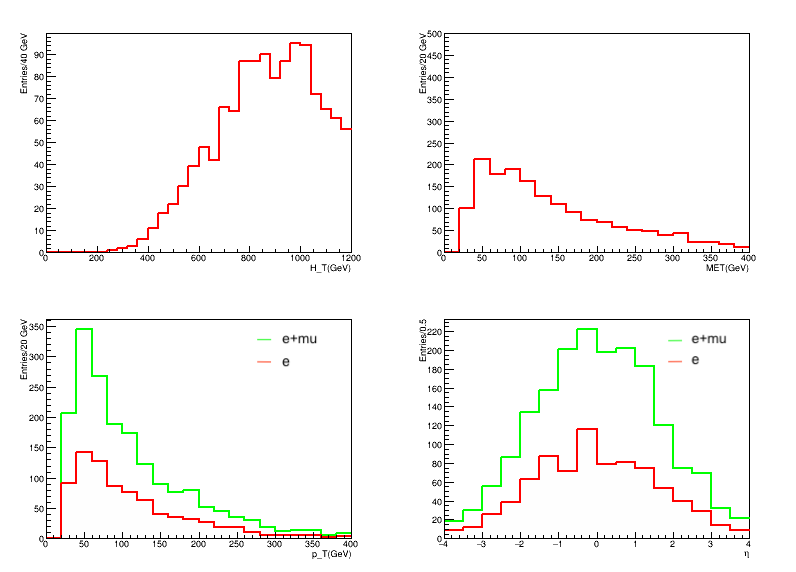}

\caption{Hadronic transverse energy ($H_{T}$) for jets, missing transverse
energy (MET) and lepton (e and e+mu) kinematical distributions ($p_{T}$
and $\eta$) for signal (with mass $m_{h^{-}}=1000$ GeV). \label{fig:fig4}}
\end{figure}

\begin{figure}
\includegraphics[scale=0.32]{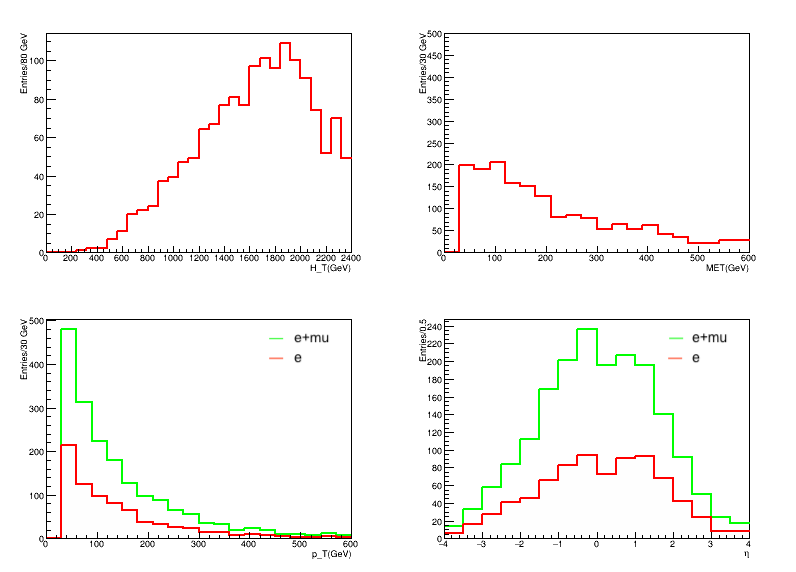}

\caption{The same as Fig. \ref{fig:fig4}, but for the mass $m_{h^{-}}=2000$
GeV. \label{fig:fig5}}
\end{figure}

Both the signal and background samples are analysed with python scripts
by reading Root trees. Events are selected as the presence of required
number of objects in the final state. We deal with events including
at least $5$ jets ($n_{jet}\geq5$) where there is at least two $b$-jets.
In addition, we require one lepton (electron or muon) and a significant
MET (focusing on $1l+MET+5jets$). At the end of the analysis histograms
are printed as figure files. The distributions of kinematical variables
($p_{T}$ of jets and leptons, $\eta$ of jets and leptons) for the
final state objects are presented in Fig. \ref{fig:fig2} and \ref{fig:fig3}
for the signal events with mass $m_{h^{-}}=1000$ GeV and $m_{h^{-}}=2000$
GeV, respectively. In Fig. \ref{fig:fig4} and \ref{fig:fig5}, the
hadronic transverse energy ($H_{T}$) for jets, missing transverse
energy (MET) and lepton (both electron (e) and electron+muon (e+mu))
kinematical distributions ($p_{T}$ and $\eta$) for signal with mass
$m_{h^{-}}=1000$ GeV and $m_{h^{-}}=2000$ GeV, respectively.

The charged Higgs boson mass is reconstructed from one top (reconstructed
from the hadronically decaying $W$ boson and subleading $b$-jet)
and the leading $b$-jet candidate. Further steps are followed as
the isolation criteria for one electron or muon (initiated from the
leptonically decaying $W$ boson), rejection of events with additional
muon or electron candidates, removal of electrons or muons if the
are separated from the nearest jet by $\Delta R<0.4$. The cut flow
for the analysis is shown in Table \ref{tab:tab3}.

\begin{table}
\caption{The cut flow for the analysis of single lepton and MET, and at least
five jets channel from charged Higgs boson associated with top quark.
\label{tab:tab3}}

\begin{tabular}{|c|c|}
\hline 
Object & Requirement\tabularnewline
\hline 
\hline 
Single electron or muon & $p_{T}>30$ GeV, $|\eta|<3.0$\tabularnewline
\hline 
Al least five jets ($n_{jet}\geq5$) & $p_{T}>30$ GeV, $|\eta|<3.0$\tabularnewline
\hline 
At least two $b$-jet ($n_{b}\geq2$) & $p_{T}>30$ GeV, $|\eta|<3.0$\tabularnewline
\hline 
Missing $p_{T}$ & $\not p_{T}>30$ GeV\tabularnewline
\hline 
($l,j$) and ($j,j$) separation & $\Delta R(l,j)>0.4$; $\Delta R(j,j)>0.4$\tabularnewline
\hline 
Hadronic transverse energy & $H_{T}>350$ GeV\tabularnewline
\hline 
Reco top mass range & $130<m_{Wb}<200$ GeV\tabularnewline
\hline 
Reco $h^{-}$ mass range & $|m_{tb}-m_{h^{-}}|<0.4m_{h^{-}}$\tabularnewline
\hline 
\end{tabular}
\end{table}

\begin{figure}
\includegraphics[scale=0.2]{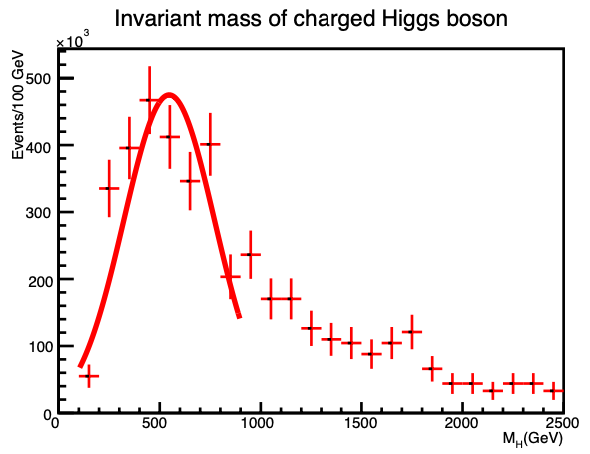}\includegraphics[scale=0.2]{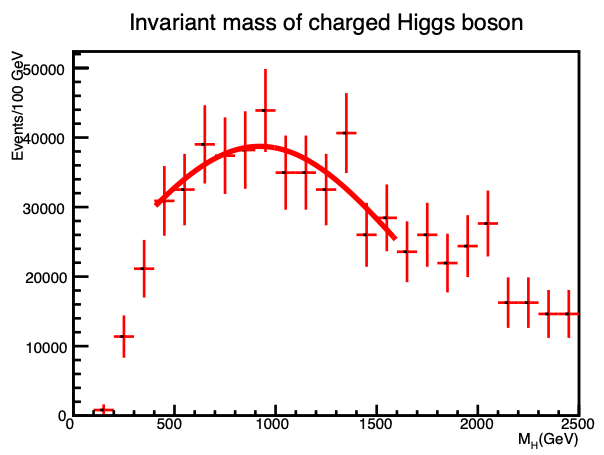}

\includegraphics[scale=0.2]{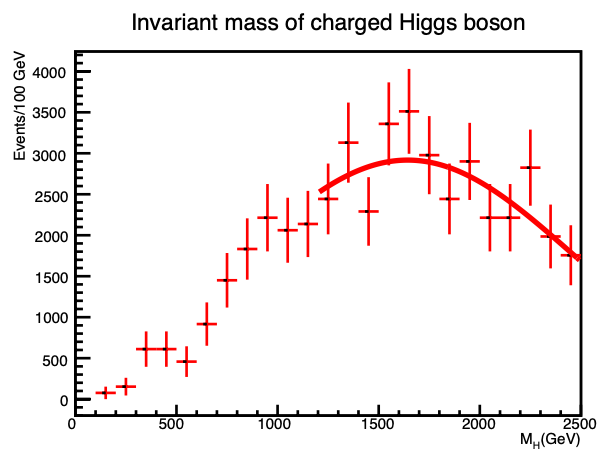}

\caption{Invariant mass distribution of four jets initiated from bottom and
top quark decays for signal with masses $m_{h^{-}}=500$ GeV, $m_{h^{-}}=1000$
GeV (upper pad) and $m_{h^{-}}=2000$ GeV (lower pad). \label{fig:fig6}}
\end{figure}

Invariant mass distribution of four jets initiated from bottom (leading
$b$-jet) and top quark are presented in Fig. \ref{fig:fig6} for
charged Higgs boson signal with masses $m_{h^{-}}=500$ GeV, $m_{h^{-}}=1000$
GeV and $m_{h^{-}}=2000$ GeV. 

We calculate statistical significance ($SS$) from signal ($N_{S}$)
and background ($N_{B}$) events within the interval $|m_{tb}-m_{h^{-}}|<0.4m_{h^{-}}$,
where significance is defined as

\[
SS=\sqrt{2\left[(N_{S}+N_{B})\ln(1+\frac{N_{S}}{N_{B}})-N_{S}\right]}
\]

Number of signal and background events and statistical significance
for the integrated luminosity of $L=1$ ab$^{-1}$ (initial) and $L=30$
ab$^{-1}$ (ultimate) at FCC-hh are given in Table \ref{tab:tab4}.

\begin{table}
\caption{Number of signal ($N_{S}$) and background ($N_{B}$) events within
the $\Delta m=\pm0.4m_{h^{-}}$ interval and statistical significance
($SS$) for the integrated luminosity of $L=1$ ab$^{-1}$ and $L=30$
ab$^{-1}$ at FCC-hh. \label{tab:tab4}}
\begin{tabular}{|c|c|c|c|c|c|c|}
\hline 
Mass (GeV) & $N_{B}$($\Delta m$) & $\tan\beta$ & $N_{S}$($\Delta m$) &  & $SS(1)$ & $SS(30)$\tabularnewline
\hline 
\multirow{4}{*}{500} & \multirow{4}{*}{368662140} & $1$ & 2851903 &  & \emph{\noun{148.34}} & 812.49\tabularnewline
\cline{3-7} \cline{4-7} \cline{5-7} \cline{6-7} \cline{7-7} 
 &  & 7 & 95340 &  & \noun{4.96} & 27.17\tabularnewline
\cline{3-7} \cline{4-7} \cline{5-7} \cline{6-7} \cline{7-7} 
 &  & 10 & 105201 &  & \noun{5.47} & 29.96\tabularnewline
\cline{3-7} \cline{4-7} \cline{5-7} \cline{6-7} \cline{7-7} 
 &  & 30 & 697536 &  & \noun{36.32} & 198.93\tabularnewline
\hline 
\multirow{4}{*}{1000} & \multirow{4}{*}{234516253} & 1 & 327598 &  & \emph{\noun{21.38}} & 117.10\tabularnewline
\cline{3-7} \cline{4-7} \cline{5-7} \cline{6-7} \cline{7-7} 
 &  & 7 & 10993 &  & \noun{0.720} & 3.94\tabularnewline
\cline{3-7} \cline{4-7} \cline{5-7} \cline{6-7} \cline{7-7} 
 &  & 10 & 12013 &  & \noun{0.780} & 4.27\tabularnewline
\cline{3-7} \cline{4-7} \cline{5-7} \cline{6-7} \cline{7-7} 
 &  & 30 & 77940 &  & \noun{5.08} & 27.82\tabularnewline
\hline 
\multirow{4}{*}{2000} & \multirow{4}{*}{61585260} & 1 & 22825 &  & \emph{\noun{2.91}} & 15.94\tabularnewline
\cline{3-7} \cline{4-7} \cline{5-7} \cline{6-7} \cline{7-7} 
 &  & 7 & 764 &  & \noun{0.097} & 0.531\tabularnewline
\cline{3-7} \cline{4-7} \cline{5-7} \cline{6-7} \cline{7-7} 
 &  & 10 & 835 &  & \noun{0.106} & 0.580\tabularnewline
\cline{3-7} \cline{4-7} \cline{5-7} \cline{6-7} \cline{7-7} 
 &  & 30 & 5316 &  & \noun{0.677} & 3.71\tabularnewline
\hline 
\end{tabular}
\end{table}

\section{Conclusion}

We have studied the charged Higgs boson (predicted by the 2HDM type-II
or MSSM) and top quark associated production in proton-proton collisions
at the FCC-hh collider. The single production of charged Higgs boson
through $pp\to h^{-}t+X$ process have been investigated in the mass
range 500 GeV to 2000 GeV using multi-jets (at least $5$ jets) final
states with one electron or muon and missing transverse momentum.
Using the relevant SM backgrounds from the lepton+jets final states,
we obtain a significant coverage of the signal parameter space and
distinguish the charged Higgs boson-top-bottom interaction for a mass
up to 2 TeV for parameter $\tan\beta=1$ and $\tan\beta=30$ at an
integrated luminosity of $30$ ab$^{-1}$. Other possible extensions
of the Higgs sector can also be searched for a wide range of parameter
space in high energy proton-proton collisions at the FCC-hh.
\begin{acknowledgments}
This work partially supported by the Turkish Atomic Energy Authority
(TAEK) under the grant No. 2018TAEK(CERN)A5.H6.F2-20. The numerical
calculations reported in this paper were partially performed at TUBITAK
ULAKBIM, High Performance and Grid Computing Center (TRUBA resources).
\end{acknowledgments}

\section*{Appendix}

The coefficients of the terms in the matrix element squared (Eq. (\ref{eq:4}))
are given 
\begin{align}
A_{1}(\hat{s},\hat{t},m_{h^{-}}) & =m_{t}^{2}\left[m_{b}^{4}+m_{b}^{2}(3m_{h^{-}}^{2}-2m_{t}^{2}-4\hat{s}-\hat{t})\right.\nonumber \\
 & +\left.\hat{s}(\hat{s}+\hat{t}-m_{h^{-}}^{2})\right]\nonumber \\
A_{2}(\hat{s}) & =-4m_{b}^{2}m{}_{t}^{2}(m{}_{b}^{2}+\hat{s})\nonumber \\
A_{3}(\hat{s},\hat{t},m_{h^{-}}) & =m_{b}^{2}\left[m_{b}^{4}+m{}_{b}^{2}(3m_{h^{-}}^{2}-2m_{t}^{2}-4\hat{s}-\hat{t})\right.\nonumber \\
 & +\left.\hat{s}(\hat{s}+\hat{t}-m_{h^{-}}^{2})\right]
\end{align}

\begin{align}
A_{4}(\hat{s},\hat{t},m_{h^{-}}) & =2m_{t}^{2}\left[m_{b}^{4}-m_{t}^{4}+m_{t}^{2}\hat{s}\right.\nonumber \\
 & +\hat{s}^{2}+m_{h^{-}}^{2}(m_{t}^{2}-\hat{s}-\hat{t})\nonumber \\
 & +m_{b}^{2}(3m_{h^{-}}^{2}-3(\hat{s}+m_{t}^{2})-\hat{t})\nonumber \\
 & +\left.(m_{t}^{2}+\hat{s})\hat{t}\right]\nonumber \\
A_{5}(\hat{t},m_{h^{-}}) & =4m_{b}^{2}m_{t}^{2}(-2m_{b}^{2}+m_{h^{-}}^{2}-2m_{t}^{2}+\hat{t})\nonumber \\
A_{6}(\hat{s},\hat{t},m_{h^{-}}) & =2m_{b}^{2}\left[m_{b}^{4}-m_{t}^{4}+m_{t}^{2}\hat{s}+\hat{s}^{2}\right.\nonumber \\
 & +\left.m_{h^{-}}^{2}(m_{t}^{2}-\hat{s}-\hat{t})\right]\nonumber \\
 & +m_{b}^{2}(3m_{h^{-}}^{2}-3(m_{t}^{2}+\hat{s})-\hat{t})\nonumber \\
 & +(m_{t}^{2}+\hat{s})\hat{t}
\end{align}

\begin{align}
A_{7}(\hat{s},\hat{t},m_{h^{-}}) & =m_{t}^{2}\left[m_{b}^{4}-2m_{t}^{4}\right.\nonumber \\
 & +m_{b}^{2}(m_{h^{-}}^{2}-4m_{t}^{2}-2\hat{s}-\hat{t})\nonumber \\
 & +\left.2m_{t}^{2}(\hat{s}+\hat{t})+\hat{s}(\hat{s}+\hat{t}-m_{h^{-}}^{2})\right]\nonumber \\
A_{8}(\hat{s},\hat{t},m_{h^{-}}) & =-4m_{b}^{2}m_{t}^{2}(m_{b}^{2}+m_{h^{-}}^{2}+2m_{t}^{2}-\hat{s}-\hat{t})\nonumber \\
A_{9}(\hat{s},\hat{t},m_{h^{-}}) & =m_{b}^{2}\left[m_{b}^{4}-2m_{t}^{4}\right.\nonumber \\
 & +m_{b}^{2}(m_{h^{-}}^{2}-4m_{t}^{2}-2\hat{s}-\hat{t})\nonumber \\
 & +\left.2m_{t}^{2}(\hat{s}+\hat{t})+\hat{s}(\hat{s}+\hat{t}-m_{h^{-}}^{2})\right]
\end{align}
when the mass of $b$-quark ($m_{b}$) is neglected, these terms reduces
to a simplified form of coefficients $A_{1}(\hat{s},\hat{t},m_{h^{-}})=m_{t}^{2}\left[\hat{s}(\hat{s}+\hat{t}-m_{h^{-}}^{2})\right]$,
$A_{4}(\hat{s},\hat{t},m_{h^{-}})=2m_{t}^{2}[-m_{t}^{4}+m_{t}^{2}\hat{s}+\hat{s}^{2}+m_{h^{-}}^{2}(m_{t}^{2}-\hat{s}-\hat{t})+\hat{t}(\hat{s}+m_{t}^{2})]$,
$A_{7}(\hat{s},\hat{t},m_{h^{-}})=m_{t}^{2}[-2m_{t}^{4}+2m_{t}^{2}(\hat{s}+\hat{t})+\hat{s}(\hat{s}+\hat{t}-m_{h^{-}}^{2})]$,
and all other coefficients vanish. In this case hadronic cross section
$\sigma(s,m_{h^{-}})$ will be proportional to $1/\tan^{2}\beta$
as mentioned in the text.

\end{document}